
\magnification=1200
\baselineskip=17pt
\vsize=22.0truecm
\voffset=0.58truecm
\hsize=5.in
\font\titlefont=cmr10 scaled\magstep2
\font\authorfont=cmr10 scaled\magstep1
\font\fromfont=cmti10 scaled\magstep0
\def\normalfrom{
\centerline{\fromfont Dipartimento di Fisica -- Universit\`a di Genova}
\centerline{\fromfont Istituto Nazionale di Fisica Nucleare --
sez. di Genova}
\centerline{\fromfont Via Dodecaneso, 33 -- 16146 Genova (Italy)}
\centerline{\fromfont E-mail: Decnet 39166; Bitnet @GENOVA.INFN.IT}
}
\def\fromhere{$^\dagger$ Dipartimento di Fisica -- Universit\`a di Genova\cr
Istituto Nazionale di Fisica Nucleare -- sez. di Genova\cr
Via Dodecaneso, 33 -- 16146 Genova (Italy)\cr
E-mail: Decnet 39166; Bitnet @GENOVA.INFN.IT\cr}
\def\ifundefined#1{\expandafter\ifx\csname#1\endcsname\relax}
\def\Month{\ifcase\month\or January\or February\or March\or April
\or May\or June\or July\or August\or September\or October\or November
\or December\fi}
\def\author{}     
\def\ABSTRACT#1{\gdef\abstract{#1}}

\def\REF#1{\gdef\ref{GEF-Th-{#1}/1992}}
\def\TITLE#1{\gdef\title{#1}}
\def\AUTHOR#1{\gdef\author{#1}}

\def\maketitle{
\vskip3.0truecm
\baselineskip=1.2\baselineskip
\ifundefined{title} \else\halign{\titlefont\centerline{##} \cr
\title\crcr}\fi
\baselineskip=\normalbaselineskip
\vskip1.2truecm
\ifundefined{author}\else\halign{\authorfont\centerline{##} \cr
\author\crcr}\fi
\vskip1.2truecm
\ifundefined{from}\normalfrom
\else\halign{\fromfont\centerline{##}\cr\fromhere\cr\from\crcr}\fi
\vfil
\ifundefined{abstract}\else{\bf ABSTRACT}\hfill\break{\abstract}\fi
\ifundefined{Note}\else\vskip1.2truecm{\bf \Note}\fi
\vfil
\hfill \ifundefined{pacs}\hfill\break\else\hfill{P.A.C.S.}\space\pacs\break\fi
\ifundefined{ref}\hfill October\space\1992\else
\ref\hfill October\space1992\fi
\eject
}

\REF{16}
\AUTHOR{Riccardo Guida and Kenichi Konishi}
\TITLE{Avalanche in the Valley\cr
 (Fermions, Anomaly and Unitarity  \cr in
 High-Energy  Electroweak Interactions)}
\ABSTRACT {Problems related to fermions, unitarity and chiral anomaly in high
energy electroweak interactions, are investigated. Particular attention
is paid to the correct functional integration over fermion fields
in the background
of instanton-anti\-instanton type configurations. This  leads to an
expansion  of  correlation functions in terms of a small parameter,
$\rho/R$, when the instanton-antiinstanton separation ($R$)
is large compared to their sizes ($\rho$).  Applying such a method to widely
discussed cases of fermion-number violation in the electroweak theory,
we  conclude that there are  no theoretical basis
for expecting  anomalous cross sections to become observable at
energies in the  $10$ TeV region.}

\maketitle
\vfill\eject

\def\dinv{{\bar D}^{-1}}
\def\et0{\eta^{(a)}_0}
\def\emi{\eta^{(i)}_m}
\def\zema{{\bar \zeta}^{(a)}_m}
\def\etm{\eta^{(a)}_m}

\def\zet0{{\bar \zeta}^{(i)}_0}

\def\zetm{{\bar \zeta}^{(i)}_m}

\def\cbar{{\bar C}}
\def\bbar{{\bar B}}
\def\d00{{\bar D}_{00}}
\def\dbar{{\bar D}}
\def\dabar{{\bar D}^{(a)}}
\def\dibar{{\bar D}^{(i)}}

\def\bra{\langle}
\def\ket{\rangle}
\def\sbar{\bar S}

In spite of much effort, the question whether instanton- (or sphaleron-)
induced electroweak interactions become strong
in  high energy scattering processes,
remains unsettled[1-20].   There are  arguments,
which use unitarity constraints on multiple gauge-boson production
processes[8,9,11,13,18],  that such processes
are always exponentially
suppressed by a finite fraction of the 't Hooft's  factor.
These arguments are quite convincing in our opinion;
however it appears difficult to make
them more quantitative.

Both  direct calculation of the cross section by instanton
method, as in the original calculation of Refs.[3,5,7], and
another approach [6,8,12,15]  which makes use
of the optical theorem and the so-called valley method,
encounter various  technical difficulties.  In some model calculations
within the second approach (see the second of Ref.[12]),
 it was claimed that the fermion number
  violating cross section approaches the geometrical one at energies of
order of the sphaleron mass.

One point in the second approach ( based on the optical theorem), however,
has always been a little obscure to us.  It seems that
 the functional integration over
 fermions
 in the instanton-antiinstanton background,
 has never been done properly.
The "fermion part of the valley trajectory" mentioned in Ref.[12]
   can easily be shown to
be inconsistent [19];  the projection operator
 introduced in Ref.[6]  was just put in by hand
and no calculation was really done with it.
  We wish to  clarify some of these and related
issues in this paper.
\smallskip
The problem to be resolved turns out to be one of quite general nature.
 Namely:
how is unitarity  obeyed in the presence of topologically nontrivial
effects such as instantons, treated within the
semiclassical approximation?\footnote{*}{The work of Ref.[10] (see also
Ref.[16]) goes quite
some way
in proving unitarity in such a situation, especially as regards
 the final states (i.e., states summed over in the
unitarity relations).  However,  fermions are not
considered there: as a result no discussion on subtleties related to
chiral anomaly is found in [10].  Note that even disregarding
fermions, a problem arises with the external (or "initial")
particles anyway.  See the last paragraph of this paper.  }

 The optical theorem states that the cross section,
$$1 + 2 \longrightarrow X, \eqno (1) $$
summed over $X$, is equal, apart from a kinematical factor, to the imaginary
part of the forward elastic amplitude,
$$ 1 + 2 \longrightarrow 1 + 2. \eqno(2) $$
Now consider a particular class of processes (1) induced by an instanton,
with the change of the fermion number,
$$\Delta f = f_1+f_2 -f_X = N_F, \eqno(3)$$
where $N_F$ is the number of the lefthanded $SU(2)$ doublets in the theory.
The sum over the final states satisfying (3) should be a part of the imaginary
part of the elastic amplitude.

For an (anti)instanton background, which is relevant for the calculation
of the production process (1), each right (left) handed fermion field
has a zero mode. The standard functional integration over fermions yields  a
product of these zero modes; by going to momentum space and by applying
the LSZ amputation one finds the S-matrix elements consistent with (3).

The corresponding contribution in the elastic amplitude (2) must arise from
a sort of instanton-antiinstanton ($i-a$)
 background, topologically (globally)
equivalent to the trivial, perturbative vacuum.  One would expect however
 that
no fermion zero modes exist in such a background (see below
for more detail). How then can one extract
the "anomalous" part of the elastic amplitude?
\smallskip
To make the problem  well-defined,
 we work with a particular class of
$i-a$ type configurations - the so called valley, or streamline, trajectory
[22,12] \footnote*{Here are some symbols and conventions used below:
 ${\bar D} \equiv D_{\mu} {\bar \sigma}_{\mu}; \,{\bar \sigma}_{\mu}
\equiv (-i, \sigma^a); $    $ a=1,2,3.$
  $ D \equiv D_{\mu} \sigma_{\mu}$,
where $\sigma_{\mu} \equiv (i, \sigma^a)$. Unless indicated by a
superscrpt $a$ (antiinstanton) or $i$ (instanton)
, the covariant derivative will refer to the valley background of Eq.
(4). The spinor indices as in $(\sigma_{\mu})_{\alpha
{\dot \alpha}}$ and  $({\bar \sigma}_{\mu})^{{\dot \alpha} \alpha}$ are
suppressed. The fermion fields $\psi_j$ are
left-handed ones.
  Throughout the paper, we work in the Euclidean spacetime: the
continuation to the Minkowski spacetime of the final formulae is
understood.   }
$$ \eqalign { A_{\mu}^{(valley)} & =
 -{i\over g}(\sigma_{\mu} {\bar \sigma_{\nu}}
-\delta_{\mu \nu} )\bigl[{(x-x_a)_{\nu} \over (x-x_a)^2 +\rho^2 } +
{(x-x_i)_{\nu} \rho^2 \over (x-x_i)^2 ((x-x_i)^2 +\rho^2) } \cr
   & + {(x-x_i +y)_{\nu} \over
(x-x_i + y )^2  }  -  {(x-x_i)_{\nu} \over
(x-x_i)^2  } ]; \cr}  \eqno(4) $$
$$  y = -R/(z-1);
\,\,   z = (R^2 + 2 \rho^2 + \sqrt{R^4 + 4\rho^2 R^2}) /2\rho^2; \,\,
R^{\mu} = (x_i - x_a)^{\mu},   $$
which is known to have the correct properties at large $R$ at least
 [7,10,12]
 to reproduce the square of the amplitude (1).
 The classical field Eq.(4)
 interpolates two solutions of the Euclidean field equations,
$A_{\mu} = A^{(i)}_{\mu} +A^{(a)}_{\mu} $ (at $R=|x_a-x_i| = \infty$)
  and  a  gauge-equivalent of
$ A_{\mu}=0 $  (at $R =0$ ).   The instanton appears
 in the  singular gauge while the anti\-instanton is in the regular
one. Moreover for simplicity we have set
$\rho_a=\rho_i=\rho$,  in Eq.(4).
 Importance of the  non-Gaussian integrations
along such an almost flat valley, was
emphasised first in Ref.[21,22] in a general context of
quantum mechanics and  QCD.

\smallskip

We wish to compute the four point function,

$$<T\psi_1(x) \psi_2(u){\bar \psi}_1(y) {\bar \psi}_2(v)>^{(A_{valley})}
$$
$$=\int{\cal D}\psi{\cal D}{\bar \psi} \, \psi_1(x) \psi_2(u) {\bar \psi}_1(y)
 {\bar \psi}_2(v)
\,{\rm e}^{-S} /Z^{(A=0)}; \eqno(5)$$
$$S=\sum_{j=1}^{N_F} \int d^4x\, i\,{\bar \psi}_j {\bar D}\psi_j      $$
in the fixed background of Eq.(4). Integrations over the collective
coordinates such as $R$ and $\rho$ are to be performed afterwards.

As the functional integral factorises in flavour we must study (suppressing
the flavour index),
$$I(x,y) = \int {\cal D}\psi{\cal D}{\bar \psi}\, \psi(x) {\bar \psi}(y)
\exp - \!\int \! d^4x\, i\,{\bar \psi}{\bar D}\psi, $$
and
$${\cal Z}=  \int {\cal D}\psi{\cal D}{\bar \psi}
\exp -\!\int\! d^4x \,i\,{\bar \psi}{\bar D}\psi = \det \dbar. $$
\bigskip
\noindent {\it Large $R/\rho$}.

At large $R$ one expects the generating functional to reduce to the product
${\cal Z}^{(a)}\cdot {\cal Z}^{(i)}$,
 where $ {\cal Z}^{(a)}$  (${\cal Z}^{(i)}$) is the generating
functional in the pure anti\-instanton (instanton) background.
To see this,
let us introduce complete sets of orthonormal modes $\{ \eta^{(a)}_n \}$
and $\{{\bar \zeta}^{(i)}_n \}$,  $n=0,1,2,....$,
  for the left-handed and right-handed
fermions, respectively.  They are eigenstates of $ D^{(a)} \dabar$
 and $\dibar D^{(i)} $:
$$  \dabar \etm = {\bar k_m} \zema\, (m=0,1,...),
  \qquad  D^{(a)} \zema = k_m \etm\,  (m=1,2,...),       $$
$$  {\bar D}^{(i)} \emi =  l_m \zetm\, (m=1,2,...),
  \qquad
D^{(i)} \zetm = {\bar l_m} \emi\,  (m=0,1,...), \eqno (6)   $$
where ${\bar k_0}={\bar l_0}=0 $.
The functional integral is then defined as:
$$\int{\cal D}\psi{\cal D}{\bar \psi} \equiv \prod_{m,n=0} da_m\,
d{\bar b}_n;  \eqno (7)$$

$$\psi = \sum_{m=0}^{\infty} a_m \eta^{(a)}_m,\qquad
{\bar \psi} = \sum_{n=0}^{\infty} {\bar b}_n {\bar \zeta}^{(i)*}_n.
$$

Now  the two point function $ I(x,y)  $ can be written as:
$$\eqalign {I(x,y) &= \det{\bar D}\, \langle x|\dinv|y \rangle\cr
      &=\det {\bar D}\, \{\, \langle x|a,0 \rangle
\langle a,0|\dinv |i,0\rangle \langle i,0|y \rangle
    + \!\sum_{m\ne 0}\langle x|a,m\rangle
 \langle a,m|\dinv|i,0\rangle \langle i,0|y \rangle\cr
     &+\sum_{n\ne 0}\langle x|a,0 \rangle
\langle a,0|\dinv|i,n\rangle \langle i,n|y \rangle
      +\!\sum_{m,n\ne 0}\langle x|a,m\rangle
\langle a,m|\dinv|i,n\rangle \langle i,n|y \rangle \}:  \cr}
\eqno(8)$$
the term proportional to the product of the zero modes
 has been singled out.  We wish to compute   Eq.(8)
 as an expansion
in the small ratio
$\,\rho/R\,$.  To do this, we first
note that the matrix $\dbar$ has the following characteristic structure.
We write
$$ \dbar = \pmatrix{d&v_1&\ldots&v_n&\ldots\cr
                         w_1&X_{11}&\ldots&X_{1n}&\ldots\cr
                 \vdots&\vdots&\ddots&\vdots&\ddots\cr
                   w_m&X_{m1}&\ldots&X_{mn}&\ldots\cr
                  \vdots&\vdots&\ddots&\vdots&\ddots\cr} \eqno(9)$$
where
$ \,d \equiv (\dbar)_{00}= \bra i,0|\dbar |a,0 \ket =\bra i,0|\cbar|a,0 \ket,
\, v_n \equiv  (\dbar)_{0n}= \bra i,0|\dbar |a,n \ket =
\bra i,0|\bbar|a,n \ket$  and
 $ w_m = \bra i,m|\dbar |a,0 \ket =\bra i,m|\cbar|a,0 \ket.$
\footnote*{
 $\cbar$ and $\bbar$ are defined as
$\dbar = \dabar +\cbar = \dibar + \bbar$, where $ \dabar$ and $\dibar$ are
the covariant derivatives in the pure antiinstanton (or instanton)
background,  and
 $ \dabar |a,0 \rangle =0;\, \bra i,0| \dibar =0$.  The
asymptotic behaviour,
$\et0 \!\!\sim \! {\rho\over (x-x_a)^3 }; \, \zet0 \!\!\sim\!
{\rho\over (x-x_i)^3 }; \,
\cbar (x)\!\!  \sim\! {\rho^2 \over (x-x_i)^3 };\, \bbar (x)\!\! \sim\!
{1 \over x-x_a }\,\, $, can be used to estimate various matrix
elements below.}

The inverse matrix $\dinv$ is given by:
$$\eqalign { (\dinv)_{00} &= 1/( d - vX^{-1}w ) \cr
 & = d^{-1} + d^{-2}v_m (X^{-1})_{mn} w_n + \cdots; \cr
(\dinv)_{mn} &= (X - d^{-1} w \otimes v) ^{-1}=
X^{-1} ( {\bf 1} - d^{-1} w \otimes v X^{-1} )^{-1} \cr
 &= (X^{-1})_{mn} + d^{-1}(X^{-1})_{ml} w_l v_k (X^{-1})_{kn}
        + \cdots,\cr
 (\dinv)_{0n} &= - d^{-1} v_l (\dinv)_{ln}, \cr
(\dinv)_{m0} &= - (\dinv)_{00} X^{-1}_{mk} w_k, \cr}  \eqno (10)$$
where  $X^{-1}$ is the inverse of the submatrix $X$.

Inserting Eq.(10) in Eq.(8)  and after some algebra  we find a simple
expression for $I(x,y)$ [19]:
$$\eqalign{ I(x,y) &= \det X\,\{\bra x|a,0\ket -
 \bra x | X^{-1}\cbar|a,0\ket \}\, \{\bra i,0|y\ket -
\bra i,0|\bbar X^{-1} |y \ket \}  \cr
 & + \det \dbar \, \bra x | X^{-1} | y \ket. \cr } \eqno(11) $$

Furthermore  $X^{-1}$  can itself be
expressed as  [19],
 $$ X^{-1} = \bigl( {\bf 1} +\sbar(\cbar - {\cal Q}_i \bbar )
\bigr)^{-1} \sbar \, {\cal P}_i, $$
where  $\sbar$ is the "propagator" in the anti-instanton background [23]:
$$ \bra x |\sbar| y \ket=\sum_{m\ne 0} \etm\!(x)\, \zema\!(y)^*/{\bar k}_m;
\quad \dabar \sbar = {\bf 1}; \quad \sbar \dabar = {\cal P}_a;$$
and
$$  {\cal P}_i \equiv {\bf 1} - {\cal Q}_i;\quad
{\cal Q}_i \equiv |i,0\ket \bra i,0|;\quad
 {\cal P}_a \equiv {\bf 1} -{\cal Q}_a;\quad
    {\cal Q}_a \equiv |a,0\ket \bra a,0|,$$
are projection operators.

Using the above expression of $X^{-1}$ in Eq.(11)
we arrive, after some reshuffling,
at  our final (and still exact) result for the two point function:
$$\eqalign{ I(x,y) &=\{\det X / (1 - f)\}\,
\bra x|({\bf 1} - {\cal G}\cbar)|a,0 \ket \, \bra i,0|({\bf 1} -
\bbar {\cal G})| y \ket \cr
 & + \det \dbar \, \bra x | {\cal G} | y \ket;  \cr
  {\cal G}  \equiv & \, \sbar \, ({\bf 1} + \cbar \sbar)^{-1}; \,\,
 f\equiv \bra i,0 | \bbar {\cal G} | i,0 \ket.  \cr} \eqno(12) $$

By expanding  ${\cal G}$  as  ${\cal G}= \sbar - \sbar \cbar
\sbar + \cdots $  we obtain the "overlap expansion"  of $I(x,y)$
in the small parameter
$ \rho/ R$.
The reason for that is that products such as
  $\cbar(z)\, \et0\!(z)$ and $\bbar(z)\, \zet0\!(z)$
with small overlapping support appear in the integrals. It is also easy to
see by using the explicit expression of the propagator $\sbar$ [23]
 that an extra power of $\sbar \cbar$ gives rise to a  further suppression.
 Some useful results are:
$$\eqalign{ d &  = \dbar_{00}= \int_z \zet0\!(z)^*  \cbar\!(z)\, \et0\!(z)
   \sim \rho^2/R^3;\cr
 (\dinv)_{00} & = (1-f)\,(d-\bra i,0|\bbar {\cal G} \cbar|a,0 \ket)^{-1}
 \sim  d^{-1};   \cr
\det \dbar & = ((\dinv)_{00})^{-1}\,\det X
 \simeq {\rm const.}\, \rho^2/R^3.\cr } \eqno(13)$$
\smallskip
Now, the "anomalous" part  of the elastic amplitude must arise from those
terms in Eq.(12)
which have the factorised dependence on $x$ and $y$,  i.e.,
terms of the form,  $\,\,\sim F(x)G(y) $.  The reason is that
 the initial fermion at $y$
must not be connected to the final fermion at $x$  by a chain
of propagators: it must be absorbed by the instanton.  In the semi-
classical approximation this  means factorisation.

All terms arising from the expansion of the first term of
$I(x,y)$ in Eq.(12)  have such a factorised structure. The non-factorised,
"non-anomalous" contribution comes from the second term.

A detailed analysis of Eq.(12)  allows us to show [19] then  that
the anomalous part of $I(x,y)$ is essentially
 dominated at  large $R/\rho$
  by
 $$I^{(anom)}(x,y) \simeq \et0\!(x)\,\zet0\!(y)^* + \cdots,   \eqno(14) $$
which originates in the first term of Eq.(8).
 Inserted in Eq.(5), it leads upon LSZ reduction  \footnote {*}{Some care
in the choice of gauge must be taken  in the LSZ
procedure; see [19].}
 to the anomalous
part of the elastic amplitude, required by unitarity.
\vfill\eject

\noindent {\it Fermions in the intermediate state}

To reach the above conclusion truely, however, we must make
one further check. For
the leading  term of Eq.(14) to represent the anomalous process,
the inter\-medi\-ate state must contain
fermions satisfying the selection rule,
Eq.(3).

That this is indeed so can be seen from Eq.(13). The functional integration
yields, for each flavour ($i=3,4,\cdots
 N_F$) a factor
$$\det \dbar / \det {\bar \partial}  \sim \rho^2/R^3, $$
but this is precisely the factor expected for a left-handed fermion,
 produced at the anti\-instanton center (with amplitude
$ \rho $ ), propagating  freely  backward to
 the instanton  (with amplitude, $
\sim 1/(x_a-x_i)^3 = 1/R^3 $ ) and absorbed by the latter (with amplitude,
 $\, \rho $ ).

To be even more explicit, suppose that we are going to observe the
intermediate  state
by setting up an appropriate detector. This would correspond to introducing
a source (or sink) term for each flavour,
$\int dx\,J^i_{\mu}(x)\psi_i(x) + h.c., $
and taking the first derivative with respect to the sources. This would
produce  pairs of zero modes (as in Eq.(14)): the fermions required by the
topological selection rule are indeed there, as long as $R/\rho$ is large.

\bigskip
\noindent{\it Small $R/\rho$. }

The overlap expansion of Eq.(12)
 fails  at small $R/\rho \le 1 $ for
obvious reasons. In particular, in the limit  $R/\rho \rightarrow 0 $,
the classical field $A_{\mu}^{(valley)} $ reduces to the trivial,
perturbative vacuum. In the free theory, the procedure adopted above is
still formally valid, but the appearance of the product of the zero modes
(as in Eq.(14)) is of course a fake, the total two-point function being
simply $\sigma_{\mu} (x-y)_{\mu}/(x-y)^4.$  This means that at
certain $R/\rho \sim O(1) $ the first term of Eq.(14), corresponding to the
anomalous process, must be effectively cancelled by contributions from other
terms (left implicit in Eq.(14)). It probably  does not make sense  to
ask exactly at which value of $R/\rho$ this occurs, but there is a fairly
good evidence that such a transition takes place rather abruptly.

We have in fact computed numerically the integral of the topological
density,
$$ C(x_4) = - \int_{-\infty}^{x_4} \int d^3x {g^2\over 16 \pi^2}
Tr (F_{\mu \nu}
 {\tilde F}_{\mu \nu})= {\cal N}_{CS}(x_4)- {\cal N}_{CS}(-\infty),
 \eqno (15) $$
as a function of  $x_4$
  for several values of $R/\rho$, for the  background of Eq.(4).
 ( ${\cal N}_{CS}$ is the Chern-Simons number.)
See Fig.1.
It is clearly seen that at  $R/\rho \ge 10 $ the topological structure
is well separated and localised at the two instanton centers, while for
small $R/\rho \le 1 $  the gauge field evidently
collapses to some insignificant fluctuation
around zero. In the latter situation it is  expected
that the level crossing [24], hence the fermion number violation, cannot
occur.

An analogous observation for the Abelian Higgs model was made
in [25].
\vfill\eject
\noindent{\it Spectrum of the Dirac operators in the valley background}

It is quite remarkable that the dominance of the fermion number violating
term (Eq.(14)) at large instanton -antiinstanton separation, occurs without
there being a single, dominant mode of the Dirac operators  $\dbar$  or
 $D$.
The result Eq.(13) for $\det \dbar$  neither implies the existence of a
particular eigenmode with eigenvalue, $\sim \rho^2/ R^3 $,  nor requires
that such
a mode should dominate over others.

Indeed, for any finite $R$ we can establish the following. First of all,
no exact zero mode satisfying $ \dbar \psi =0 $ or $ D {\bar \psi} =0$
exists. (This can be shown by a direct calculation [19].)
  Secondly, there are many non-zero modes, definitely lying below
  $\rho^2/ R^3$. They form a continuous spectrum,
reaching down to $0$,  if we work
in the whole of ${\bf R}^4$.  In
particular, for the lowest lying  modes with  $k \ll   \rho^2/ R^3, $
 both
the eigenvalue and the wave function differ little  from the free
spectrum.

This last point can be made more explicit by putting the system in a large
box of size $L^4$ such that
$$ L \gg R, \rho   \eqno(16)$$
 and by applying the
standard perturbation theory, treating the valley background of Eq.(4) as
a perturbation.  We may conveniently  consider
 the Hermitian "Hamiltonian"
$\dbar D$  or $ D \dbar $.  For the lowest-lying levels and to first order,
we find:
$$  E^{(0)} = {\bf n}^2(\pi/L)^2,
\quad ({\bf n}=(n_1,n_2,n_3,n_4);\,\, n_j=1,2,..); $$
$$ |E^{(1)}| \le const. {\rho \over  L^3}\ll
E^{(0)}.    \eqno(17)$$
 It can be readily verified that the wave functions are also modified only
by a small amount.

(In passing, this shows that a single $i-a$ pair in itself cannot lead to
chiral symmetry breaking  in QCD: the latter
necessarily  requires [26]
an accumulation of  eigenvalues of $D, \dbar$  towards $0$.
In the context of
instanton physics,  that would require something like
the  "instanton liquid" (Ref.[27]).)

The results such as Eq.(13) and Eq.(14) are thus  collective
effects in which many modes contribute together.  No single mode plays any
particular role.

\bigskip
\noindent {\it An apparent paradox and its resolution.}

We then seem to face a paradox.  At large instanton
 antiinstanton separation, physics must factorise in certain sense, and
we do find results (e.g., Eq.(14) )
consistent with such intuition. This is fine. The problem is that
the mathematics to achieve this looks
very different from that of the usual instanton physics where a single
fermion zero mode plays a special role. In our case, there is no hint even of
the presence of a quasi zero mode (Eq.(17)).
What is going on?

The key to the resolution of this apparent paradox is the
inequality, (16). In order to be able to compute the S-matrix
elements from the four point function (LSZ procedure),
we are indeed forced to work in a spacetime region whose linear size
($L$)  is much greater than
both of the physical parameters $R$ and $\rho$,
 independently of the ratio $R/\rho$.
  All
fields must be normalised in such a box with an appropriate boundary
condition.  It is clear that this system (with the  $\,i-a \,$
background)
cannot be connected smoothly  to the system with a single
(say)  antiinstanton inside the box.  The gauge field topology
remains firmly in the trivial sector.

In spite of our use of the "zero modes" $\et0\!,\, \zet0$  as a convenient
device for calculation,  they are  not
good approximate wave functions for the lowest modes in the valley
background, however large $R/\rho$ might be.
In fact, had one insisted upon using $\et0$ as the "unperturbed" wave
function,
one would have discovered that the effect of $\cbar$  was always
non\-perturbative and large (near $x_i$). \footnote*{ The standard
perturbation
theory applied to  $H = D \dbar = D^{(a)} \dabar \,  + H^{\prime}$
with $H^{\prime}=D^{(a)}\cbar + C \dabar + C\cbar$,
yields  $\Delta E =0 $ to all orders,
 reflecting the topological stability of the
fermion zero mode. (Another way to see this is to notice a
supersymmetric structure underlying the system. We thank C.Imbimbo for
pointing out this to us.)
Such a result, however, is false in the case of topol\-ogy-
changing modification of the gauge field. }

All this is  to be distinguished clearly from the situation where we ask
e.g., what the effects of an instanton on a distant planet are.
  Such a case would
correspond to the inequality opposite to (16)
$$ R \gg L, $$
if we restrict ourselves to the instanton-antiinstanton case.  As long as
we are interested in physics inside our laboratory (or on the earth, anyway)
both physics and mathematics are
described by just ignoring the distant instanton,
to a good approximation.  (Particles are produced and detected inside a
box of volume, $L^4 \ll R^4$,
 the gauge field having the winding number $-1$,
fields and functional integrations defined in the same box, etc.)
The effect of the distant instanton is a true, and negligibly small,
perturbation in this case.

\bigskip

\noindent {\it Fermion number violation
in high energy scattering
 processes.}

As is well-known,
 taking into account the
contribution of the Higgs field in the action and using the saddle point
approximation in the integrations over the instanton
collective coordinates, one finds
a relation among the c.m. energy $\sqrt s $ and the saddle point
values of the parameters $R$ and $\rho$. In a simple toy-model calculation
(the second of of Ref.[12]) which uses the valley trajectory of Eq.(4),
 it was found that at an energy of the order of the sphaleron
mass the total action  (at the saddle point) vanishes
and that the 't Hooft suppression factor disappears.

But at that point,  the corresponding saddle point values of the
instanton parameters are found to be   $ \rho = 0;\, R/\rho = 0 $ (hence
$A_{\mu} = 0 $) : the unsuppressed cross section should simply correspond
to a non-anomalous, perturbative cross section. This was pointed out  by
some authors [13] also.

What happens is that the anomalous term of Eq.(14) rapidly
disappears at $R/\rho \sim 1 $ as discussed above,
precisely where the valley action sharply
drops to zero [15]. We must conclude that the
toy-model considered there
shows no sign of anomalous cross section (associated with the
production of large number ($\sim 1/\alpha $) of gauge and Higgs bosons)
becoming large, contrary to the original claim.

Of course, there is no proof that all sorts of quantum correct\-ions to the
inst\-anton-induced process (1) are correctly described by a  classical
background such as Eq.(4). But then there are as yet no calculations anyway
which show that the anomalous process becomes observable in high energy
scatterings.

More generally, the results of this paper imply that, in order for the
fermion-number violating cross sections to become observable at high
energies, a new mechanism must be found in which the
background field governing the elastic amplitude Eq.(2)
does not effectively reduce to $ A^{eff}_{\mu} = 0 $.  Note that this is
necessary, whether or not multiple-instanton type configurations become
important at some energies. It is  difficult to envisage
 such a novel mechanism, not accompanied by some finite fraction of the
't Hooft factor.
\bigskip
\noindent {\it Theories with no fermions.}

It is often stated that fermions are not essential,  as dynamical
effects  induced by inst\-ant\-ons  can well be studied in a theory
without fermions.
This is certainly true, but it does not mean
that the consideration of this paper is
irrelevant in such a case.

Quite the contrary.  The crucial factor,
$$\exp(ik_j\cdot x_a) \exp(-ik_j\cdot x_i) \qquad (j=1,2) $$
associated with the external particles,
appears upon LSZ amputation applied to the product
of the zero modes of Eq.(14)  in the case of processes
 with fermions.  In the case of external gauge bosons,
the same factor emerges as a result of the "semiclassical" approximation,
$$\eqalign {& <TA_{\mu}(x) A_{\nu}(u) A_{\rho}(y) A_{\sigma}(v)> \cr
& \simeq <TA^{(a)}_{\mu}(x) A^{(a)}_{\nu}(u) A^{(i)}_{\rho}(y)
 A^{(i)}_{\sigma}(v)>  + .... \cr }\eqno(18) $$
It is however clear  that this approximation, which is good at
large $R/\rho$, fails at  $R/\rho < 1 $  just as Eq.(14) does.
The first term of Eq.(20), the anomalous term,  disappears
 precisely when the
cross section is claimed to become observable.
\bigskip

\noindent {\bf Acknowledgments}

Discussions with
 D. Amati, C. Imbimbo, M. Maggiore, H. Leutwyler and
V. Zakha\-rov, have been especially helpful.
We  are also grateful to many other collegues, M.Bertero,
 U. Bruzzo, M. Mint\-chev and
M. Por\-rati among them,   for useful information.
   Carlo Musso has collaborated
at an early
stage of the work.   Part of this work was done during a
visit to the Theory Division of CERN.    One of us (K.K.) thanks
the Physics Department of University of Pisa for hospitality.
\bigskip
\noindent {\bf References}

\smallskip
\item {[1]} N. Manton, Phys.Rev. {\bf D28} (1983) 2019.
\item {[2]} H. Aoyama and H. Goldberg, Phys. Lett. {\bf B188} (1987) 506.
\item {[3]} A. Ringwald, Nucl.Phys. {\bf B330} (1990) 1;
       O. Espinosa, Nucl.Phys. {\bf B343} (1990) 310.
\item {[4]} J. Cornwall, Phys.Lett. {\bf B243} (1990) 271.
\item {[5]} L. McLerran, A. Vainshtein and M. Voloshin,
    Phys. Rev. {\bf D42} (1990) 180;
       P. Arnold and L. McLerran,
  Phys. Rev. {\bf D37} (1988) 1020.
\item {[6]} M. Porrati, Nucl.Phys. {\bf B347} (1990) 371.
\item {[7]} S. Khlebnikov, V. Rubakov and P. Tinyakov,
  Nucl.Phys. {\bf B350} (1991) 441.
\item {[8]} V. Zakharov, Nucl. Phys. {\bf B371} (1992) 637;
   Nucl. Phys. {\bf B353} (1991) 683; Munich preprint,
 MPI-PAE/\-PTh 11/91(1991).
\item {[9]} G. Veneziano, unpublished note (1990);  CERN TH-6399/92 (1992).
\item {[10]} P. Arnold and M. Mattis, Phys.Rev. {\bf D42} (1990) 1738;
         Phys.Rev.Lett. {\bf 66} (1991) 13; Phys. Rev. {\bf D44} (1991) 3650.
\item {[11]} A. Mueller,
    Nucl.Phys. {\bf B353} (1991) 44; Colum\-bia preprint CU-TP-548 (1992).
\item {[12]} V.V. Khoze and A. Ringwald, Nucl.Phys. {\bf B355} (1991) 351;
       Phys.Lett. {\bf 259} (1991) 106; CERN-TH-6082/91.
\item {[13]} M. Maggiore and M. Shifman, Nucl. Phys. {\bf B365} (1991)161;
Nucl. Phys. {\bf B371} (1992) 177;
 TPI-MINN-92/2-T (1992),  to appear in Phys. Rev. D.
\item {[14]} S.Yu. Khlebnikov, V.A. Rubakov and P.G. Tinyakov,
Nucl. Phys. {\bf B350} (1991) 441.
\item {[15]} V.V. Khoze, J. Kripfganz and A. Ringwald, Phys. Lett.
{\bf B275} (1992)381; {\bf B277} (1992) 496; {\bf B279} (1992) 429E.
\item {[16]} O.R. Espinosa, Univ. Washington preprint, UW/PT-91-12 (1991).
\item {[17]} D. Dyakonov and V. Petrov, Phys. Lett. {\bf B275} (1992) 459.
\item {[18]} K. Konishi and C. Musso, Univ. of Genoa preprint, GEF-TH-14 (
1991).
\item {[19]} R. Guida and K. Konishi, in preparation.
\item {[20]} M. Voloshin, Nucl. Phys. {\bf B363} (1991) 425.
\item {[21]} I. Balitsky and A. Yung, Phys. Lett. {\bf B168} (1986) 113.
\item {[22]} A. Yung, Nucl. Phys. {\bf B297} (1988) 47.
\item {[23]} L. Brown, R. Creamer, D. Carlitz and C. Lee, Phys. Rev. {\bf
D17} (1978) 1583;  D. Amati, G.C. Rossi and G. Veneziano,
Nucl. Phys. {\bf B249} (1985) 1.
\item {[24]} C.G. Callan, R. Dashen and D.J. Gross, Phys. Rev. {\bf D17}
 (1978) 2717.
\item {[25]} M.Shifman, A seminar at
 Yale (private communication by M.Maggiore).
\item {[26]} H. Leutwyler, Bern preprint, BUTP-91\--43 (1991).
\item {[27]} E.V. Shuryak, Nucl. Phys. {\bf B341} (1990) 1; D.I. Dyakonov
and V.Yu. Petrov, Nucl. Phys. {\bf B272} (1986) 457.

\bigskip
\noindent {\bf Figure Caption}
\smallskip
\noindent Fig.1 $C(x_4)$ of Eq.(15) versus $x_4/\rho$
for  $R/\rho=10$ (outmost curve),
 $R/\rho=5,$   $R/\rho=2$  (middle),   $R/\rho=1,$ and
   $R/\rho=0.5$ (innermost curve).   The instanton and antiinstanton are
situated at $ ({\bf 0}, R/2) $ and at $ ({\bf 0}, -R/2) $, respectively.
\vfill\eject
\end